\documentclass[12pt,preprint]{aastex}

\newcommand{\myemail}{levenson@physics.ucla.edu}

\shorttitle{DIRBE minus 2MASS}
\shortauthors{Levenson et al.}

\begin{document}

\title{DIRBE Minus 2MASS: Confirming the CIRB in 40 New Regions at 2.2 and 3.5 Microns}

\author{L.R. Levenson \& E.L. Wright }
\affil{Department of Physics and Astronomy, University of California, Los Angeles, CA 90095-1562}
\email{\myemail}

\author{B.D. Johnson}
\affil{Dept. of Astronomy, Columbia University, 550 W. 120th St., New York, NY 10027}

\begin{abstract}
With the release of the 2MASS All-Sky Point Source Catalog, stellar fluxes from 2MASS are used to remove the contribution due to Galactic stars from the intensity measured by DIRBE in 40 new regions in the North and South Galactic polar caps.  After subtracting the interplanetary and Galactic foregrounds, a consistent residual intensity of  14.69 $\pm$ 4.49 kJy sr$^{-1}$ at 2.2 $\mu$m is found.  Allowing for a constant calibration factor between the DIRBE 3.5 $\mu$m and the 2MASS 2.2 $\mu$m fluxes, a similar analysis leaves a residual intensity of 15.62 $\pm$ 3.34 kJy sr$^{-1}$ at 3.5 $\mu$m.  The intercepts of the DIRBE minus 2MASS correlation at 1.25 $\mu$m show more scatter and are a smaller fraction of the foreground, leading to a still weak limit on the CIRB of 8.88 $\pm$ 6.26 kJy sr$^{-1}$ (1 $\sigma$).   

\end{abstract}


\section{Introduction}

The Cosmic InfraRed Background (CIRB) is the aggregate of the short wavelength radiation from the era of structure formation following the decoupling of matter and radiation in the early universe, redshifted to near infrared wavelengths by cosmological redshifting and far infrared wavelengths by dust-reprocessing, i.e. absorption and re-emission of starlight by intervening dust. Particle decay models also allow for a contribution to the CIRB from photons produced in the decay of weakly interacting massive particles such as big bang relic neutrinos \citep{bond86}. The Diffuse InfraRed Background Experiment (DIRBE) on the COsmic Background Explorer (COBE, see \citet{bog92}), which observed the entire sky in 10 infrared wavelengths from 1.25 to 240 $\mu$m, was primarily intended to measure the CIRB.  A determination of the CIRB turned out to be exceedingly difficult due to bright foreground contamination including Galactic starlight and the strong thermal emission and reflection of sunlight by dust in the solar neighborhood in many of the DIRBE bands, especially in the near infrared (NIR).  The DIRBE team estimated this zodiacal contribution by fitting a computed brightness, based on a physical model of the InterPlanetary Dust (IPD) cloud, to the weekly variations in the DIRBE brightness due to COBE's changing position in the solar system as the Earth orbited the Sun \citep{kel98}.  The modeled brightness was computed by integrating a source function, multiplied by a three 
dimensional density distribution function, along the line of sight. Using this model for zodiacal light subtraction, the DIRBE team succeeded in measuring an isotropic CIRB of 25 $\pm$ 7 nW m$^{-2}$ sr$^{-1}$ (1170 $\pm$ 328 kJy sr$^{-1}$) and 14 $\pm$ 3 nW m$^{-2}$ sr$^{-1}$ (1123 $\pm$ 241 kJy sr$^{-1}$) at 140 and 240 $\mu$m respectively \citep{hau98}.  In the NIR bands, they reported only upper limits on the CIRB. 

Using the ``very strong no-zodi principle'' of \citet{wr97}, another model of the 
zodiacal light was determined and is described in \citet{wr98} and \citet{gor00}.  After creating DIRBE all sky maps with this new model subtracted, the problem of subtracting Galactic starlight in the NIR was addressed by \citet{gor00} at 2.2 and 3.5 $\mu$m by directly observing Galactic stars in a 
$2^\circ \times 2^\circ$ region of the sky, smoothing the observed intensities to the DIRBE pixel size and 
subtracting the resulting intensities from DIRBE maps with the \citet{wr98} zodiacal light model subtracted. \citet{elw01} used this same procedure of subtracting Galactic starlight from the DIRBE maps using stellar fluxes from the 2MASS 2nd incremental release Point Source Catalog (PSC) in four regions in the North and South Galactic poles.  \citet{dwe98} showed that the DIRBE 3.5 $\mu$m intensity was strongly correlated with the DIRBE 2.2 $\mu$m intensity at high Galactic latitudes and used this correlation to convert a lower limit on the 2.2 $\mu$m CIRB, based on galaxy counts, into a lower limit on the 3.5 $\mu$m CIRB.  \citet{wrj01} extended the \citet{elw01} analysis to 13 regions at 2.2 $\mu$m and combined the \citet{elw01} and \citet{dwe98} techniques to obtain estimates of the CIRB for those 13 regions at 3.5 $\mu$m.  In this work we extend the \citet{wrj01} analysis to 40 new regions to further constrain the CIRB at 1.25, 2.2 and 3.5 $\mu$m.

\section{Data Sets}

The two main datasets in this paper are the zodi-subtacted DIRBE maps and the 2MASS All-Sky Point Source Catalog (PSC).

DIRBE has a large $0.7^\circ \times 0.7^\circ$ square beam with a diagonal of $1^\circ$.  Pixel intensities in the DIRBE maps are averages of all observations made while the beam was centered in a given pixel in the COBE Quadrilateralized Spherical Cube (CSC) projection.  Due to the large beam size, bright stars outside of a particular pixel will, depending on the exact center position and position angle of the beam, occasionally affect the observed brightness in that pixel.  Thus a thick buffer ring is needed around any studied field to keep bright stars outside the field from influencing the measured DIRBE intensity.  The resulting inefficiency was minimized by using the darkest possible $2^\circ$ circular regions which have the largest possible area:perimeter ratio.  A list of dark spots was created by smoothing the 3.5 $\mu$m DIRBE Zodiacal Subtracted Mission Average (ZSMA) map to 1/64th it's original resolution, reducing the 393,216 $0.32^\circ \times 0.32^\circ$ pixels to 6144 pixels approximately $2.5^\circ \times 2.5^\circ$ using a straight average of nearest neighbor pixels and sorting the resulting low resolution map.  The 40 darkest regions not used in previous analyses were used.  

However, the DIRBE ZSMA project data set only uses a fraction of the DIRBE data since extreme solar elongations were dropped.  Therefore, a set of mission averaged zodiacal subtracted DIRBE maps were created by \citet{elw01} using the zodiacal light model of \citet{wr98}.  The physical model is similar to the \citet{kel98} model, with an added constraint.  The ``very strong no-zodi principle,'' described in \citet{wr97} and \citet{gor00}, requires that at high Galactic latitudes, at 25 $\mu$m, the background emission should be isotropic and adds a single pseudo-observation of zero emission to the nearly $10^5$ observations used to fit the time variation in the weekly maps to the model brightness.  At 1.25 and 2.2 $\mu$m, no correction for interstellar dust emission is needed, while at 3.5 $\mu$m, there is a very small correction \citep{are98}.  The $i^{th}$ pixel in these maps provides the DIRBE data $DZ_i$.

Using the NASA/IPAC InfraRed Science Archive (IRSA), fluxes from all stars in each of the 40 regions brighter than K = 14 in the all-sky release of the 2MASS PSC were obtained.  Again due to the large beam size and effectively random distribution of the beam center and position angle, these fluxes were converted into intensities by smearing with a $0.7^\circ \times 0.7^\circ$ square beam with a center uniformly distributed in the DIRBE pixel and orientation uniformly distributed in position angle, to obtain the cataloged star contribution, $B_i$, to the DIRBE intensity.  This smearing process is described fully in Section~\ref{analysis}.  For the J-band contribution, only stars with J and K less than 14 were used.  This dual wavelength magnitude selection is essentially equivalent to a simple $J <$ 14 selection \citep{elw01}.  

\section{Analysis}
\label{analysis}

\begin{figure}
\epsscale{.80}
\plotone{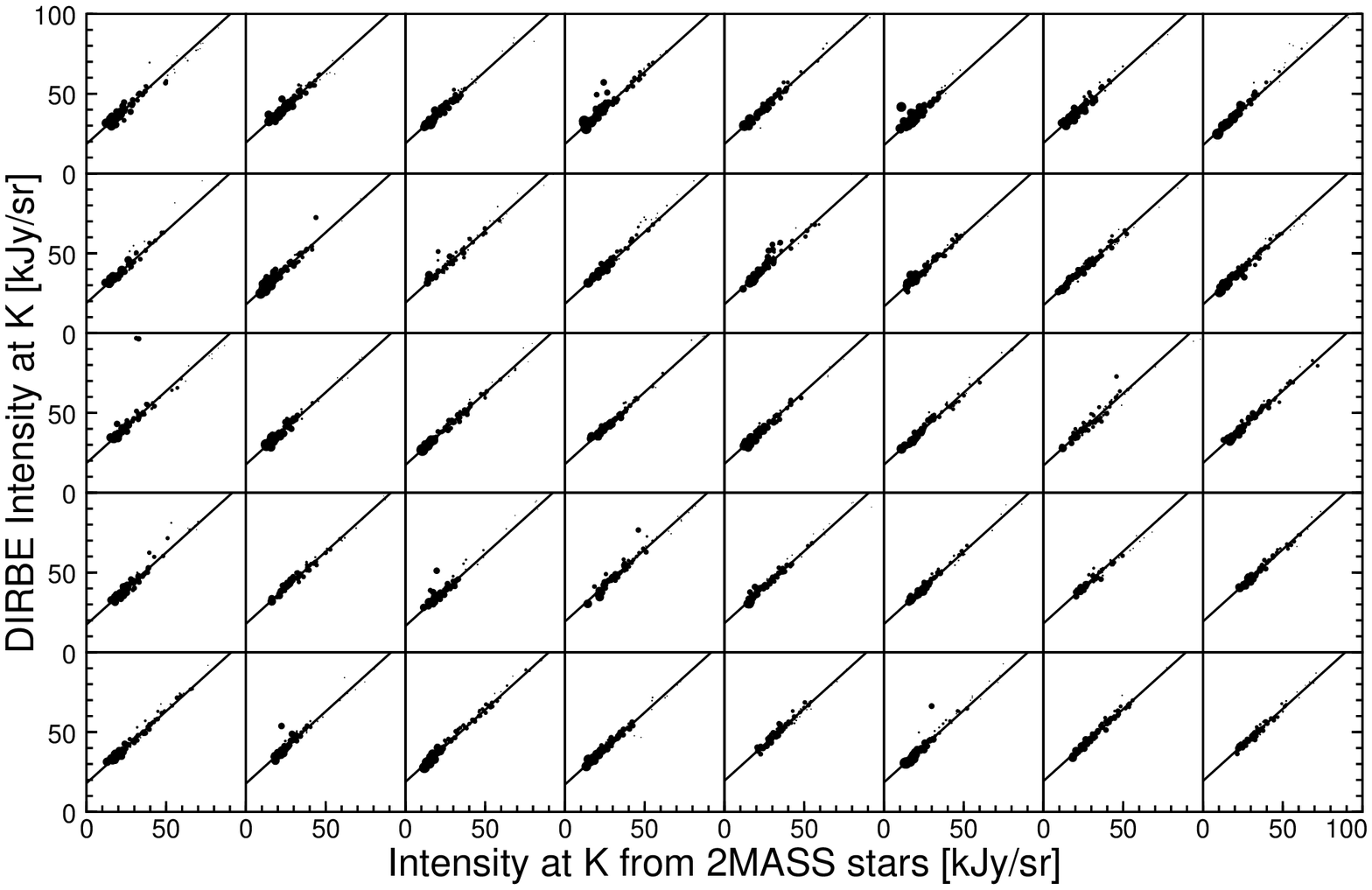}
\caption{K-band: DIRBE 2.2 $\mu$m intensities vs. 2MASS 2.2 $\mu$m stellar intensities for the forty new regions.  Fitted lines show a weighted median fit, resulting in  a common slope of 0.88 and 40 different intercepts, to the data points.  Intercepts along with region l,b and $\beta$ are listed in Table~\ref{ktable}. Reading left to right and top to bottom, the panels are in the same order as Table~\ref{ktable}. Point sizes are inversely proportional to $\sigma_i$.} 
\label{kbdz}
\end{figure}

\begin{figure}
\epsscale{.80}
\plotone{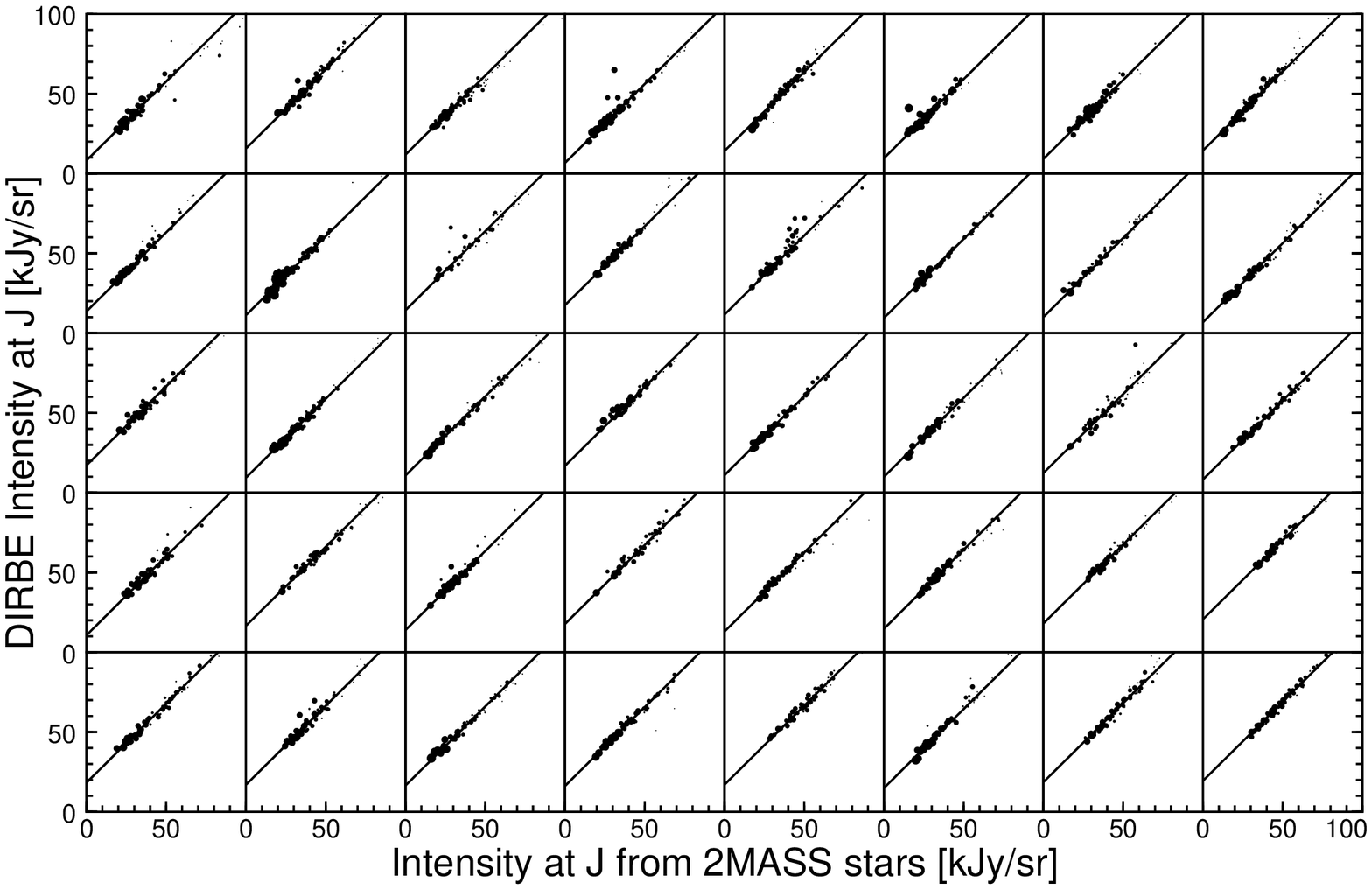}
\caption{J-band:  DIRBE 1.25 $\mu$m Intensities vs. 2MASS 1.25 $\mu$m stellar intensities for the forty new regions.  Fitted lines show a weighted median fit, resulting in  a common slope of 0.97 and 40 different intercepts, to the data points.  Intercepts along with region l,b and $\beta$ are listed in Table ~\ref{jtable}.  Reading left to right and top to bottom, the panels are in the same order as Table~\ref{jtable}.  Point sizes are inversely proportional to $\sigma_i$.} 
\label{jbdz}
\end{figure}

\begin{figure}
\epsscale{.80}
\plotone{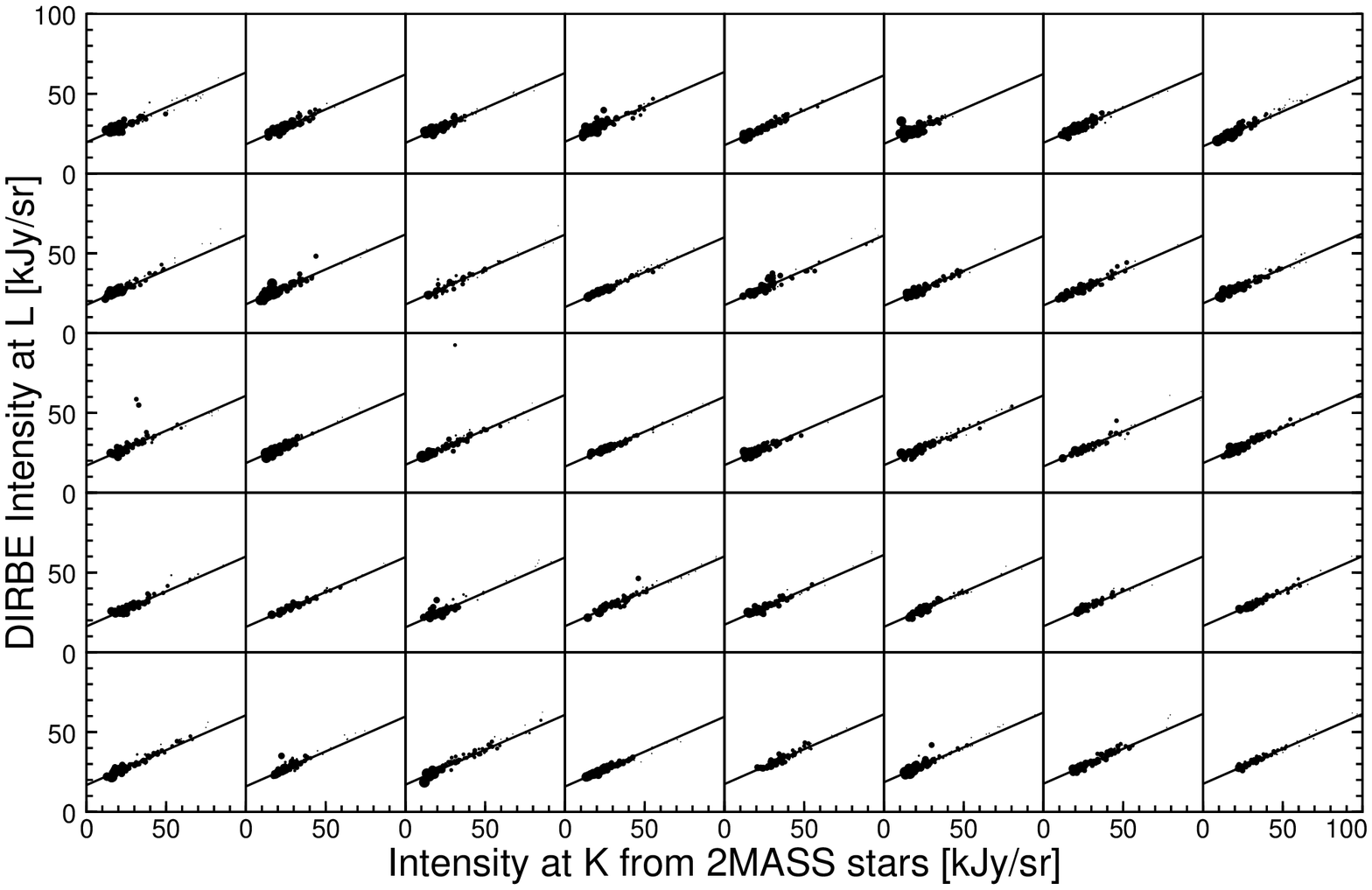}
\caption{DIRBE L vs. 2MASS K:  DIRBE 3.5 $\mu$m Intensities vs. 2MASS 2.2 $\mu$m stellar intensities for the forty new regions.  Fitted lines show a weighted median fit, resulting in a common slope of 0.43 and 40 different intercepts, to the data points.  Intercepts along with region l,b and $\beta$ are listed in Table~\ref{ltable}.  Reading left to right and top to bottom, the panels are in the same order as Table~\ref{ltable}.  Point sizes are inversely proportional to $\sigma_i$.} 
\label{lbdz}
\end{figure}

The zodi-subtracted DIRBE intensity in the $i^{th}$ pixel, $DZ_i$, should be the sum of the cataloged stars, $B_i$; the faint stars, $F_q$, assumed constant for all pixels in the $q^{th}$ region; and the CIRB, C ,which is isotropic, i.e.   
\begin{equation}
DZ_i = B_i + F_q + C \, .
\end{equation}
The cataloged star contribution was computed using the ``smearing'' described in \citet{gor00} and \citet{elw01},
\begin{equation}
B_i = \frac{1}{{\Omega_b}}\sum_j p_{ij} S_j \, ,
\end{equation}
where $\Omega_b$ is the solid angle of the DIRBE beam, $S_j$ is the flux of the $j^{th}$ star and $p_{ij}$ is the probability of the $j^{th}$ star affecting the $i^{th}$ pixel under the assumptions that over the many observations in which the DIRBE beam was centered in the $i^{th}$ pixel, the beam center was uniformly distributed within the pixel and the beam orientation is uniformly distributed in position angle.  Due to the random distribution of the orientation of the square $0.7^\circ \times 0.7^\circ$ beam, the probablility as a function of the angular distance, r[($\alpha, \delta), (\alpha_\star, \delta_\star$)], between the beam center ($\alpha, \delta$) and a particular star at ($\alpha_\star, \delta_\star$) is 
\begin{equation}
P(r) = \cases{0, &for $r > \frac{l}{\sqrt{2}}$;\cr 1-\frac{4}{\pi} \arccos{\frac{l}{2r}}, &for $\frac{l}{2} \leq r \leq \frac{l}{\sqrt{2}}$;\cr 1, &for $r < \frac{l}{2}$;\cr}
\end{equation}
where $l$ is the width of the beam (0.7 deg) and, 
\begin{equation}
r = \arccos[\cos(\alpha - \alpha_\star)\cos(\delta)\cos(\delta_\star) + \sin(\delta)\sin(\delta_\star)] \, .
\end{equation}
Then, to account for the random position of the beam center within the pixel, this probability must be averaged over the area of the pixel by integrating P(r) over the solid angle of the pixel and dividing by the pixel solid angle, $\Omega_i$, so that  
\begin{equation}
p_{ij} = \frac{\int_{\Omega_i}P(r)d\Omega_i}{\Omega_i} \, .
\end{equation}
Uncertainties in $B_i$ were calculated as in  \citet{elw01}: 
\begin{equation}
\sigma^2(B_i) = \frac{1}{{\Omega_b}^2}\sum_j[p_{ij}(1-p_{ij})+{p_{ij}}^2(0.001+(0.4\ln 10)^2\sigma^2(m_j))]{S_j}^2 \, .
\end{equation}
The first term is noise due to stars near the edge of the DIRBE beam which will``flicker'' in and out of the beam when observed with various centers and position angles.  The second term is the flux error with an added allowance for variation in the fluxes between the DIRBE and 2MASS observations as in \citet{elw01}, with the modification that the allowance for variable stars was reduced from 0.1 to 0.001, reducing this allowance from $\sigma$ = 0.34 to $\sigma$ = 0.03 magnitudes.  Statistically, the $\sigma$'s computed with the \citet{elw01} value of 0.1 were too large to be justifiable.  Upon dividing the residuals from the linear fit to the DIRBE vs. 2MASS intensities (described below) by the computed $\sigma$'s, we noticed that the residual/$\sigma$ at all pixels was less than one.  The new value of 0.001 gives a statistically more reasonable distribution of residual/$\sigma$ values which is described at the end of this Section.  The error estimates for all stars brighter than K = 5.5 or J = 6.5 were set to $\pm$ 1 mag to effectively remove pixels affected by confusion due to saturation from the final analysis.  Stars with reported null uncertainties were assigned an uncertainty of $\pm$ 0.5 mag.  Since the error in the DIRBE data is negligible \citep{hau98}, all of the error comes from the calculation of $B_{i}$ and is ascribed to $DZ_{i}$ for the fits.  Figures \ref{kbdz}, \ref{jbdz} and \ref{lbdz} show plots of $DZ_i$ vs. $B_i$ for all pixels in each of the 40 regions in K, J, and L respectively where the point sizes are inversely proportional to the above $\sigma$'s.  

The fits in K, J and L have slopes of $\kappa_K$ = 0.88, $\kappa_J$ = 0.97 and $\kappa_L$ = 0.43 respectively with 40 independent intercepts in each band, derived using a weighted median procedure, i.e. finding the values of $\kappa$ and DZ(0) that minimize the sum:
\begin{equation}
\sum_{i} |(DZ_i-\kappa B_i-DZ(0))/\sigma_i| \, .
\end{equation}  
Derived intercepts for each field are given in Tables \ref{ktable}, \ref{jtable} and \ref{ltable} for K, J, and L respectively.

The contributions from stars fainter than the 14th magnitude were evaluated statistically by fitting a power series of the form $n_q$(m) = $n_{\circ,q} 10^{\alpha m}$ to counts of 2MASS stars in each of the 40 regions, binned into 3 one-magnitude bins centered on m= 11.5, 12.5 and 13.5.  The fits resulted in 40 individual $n_{\circ,q}$ and $\alpha_K = 0.288$ and $\alpha_J = 0.276$ where any $\alpha < 0.4$ results in a converging flux contribution.  The intensity contribution from faint stars in the $q^{th}$ region with solid angle $\Omega_q$ is then
\begin{equation}
F_q = \frac{F_\circ(\lambda) n_{\circ,q} }{\Omega_q} \stackrel{\infty}{\sum_{m=14.5}} 10^{(\alpha - 0.4)m}
\end{equation}
or, in the limit of infinitely fine bins,
\begin{equation}
F_q = \frac{F_\circ(\lambda) n_{\circ,q} }{\Omega_q} \stackrel{\infty}{\int_{14.5}} 10^{(\alpha - 0.4)m} dm \, ,
\end{equation}
which was computed analytically.  At L, the faint source contribution in each region is the 2.2 $\mu$m value multiplied by the calibration ratio of 0.491.  These F$_q$ are listed in Tables \ref{ktable}, \ref{jtable} and \ref{ltable} for K, J and L.  An uncertainty of 20\% of the total prediction is assigned to this correction, and is listed in Table~\ref{error} under ``Faint Source.''

The CIRB in each region is then the derived intercept, DZ(0), minus the faint star contribution F$_q$, C = DZ(0) - F. These values are also listed in Tables \ref{ktable}, \ref{jtable} and \ref{ltable}. The mean of these CIRB estimates are 14.59 $\pm$ 0.14, 8.83 $\pm$ 0.51 and 15.57 $\pm$ 0.20 kJy sr$^{-1}$ for K, J and L.  These standard deviations of the means are listed in Table~\ref{error} as ``Scatter.'' 

2MASS magnitudes at K and J were converted into fluxes using F$_\circ$(K) = 614 Jy and F$_\circ$(J) = 1512 Jy  which were derived by \citet{elw01} and \citet{gor00}.  However, the derived slopes at K and J indicate DIRBE fluxes for a zero magnitude 2MASS star of 540 and 1467 Jy respectively.  The ratio of the calibration factors at 3.5 and 2.2 $\mu$m is 0.491, consistent with those found by \citet{dwe98}, \citet{wrr00} and \citet{wrj01}.  Uncertainties in the CIRB due to calibration error  were estimated using the change in the median DZ(0) when the slopes, $\kappa$, were forced to change by $\pm$5\% at J and L, or $\pm$10\% at K due to the large difference \citep{wrj01} between the fitted value of 0.88 and the expected value of 1.  The change in the medians are $\mp$ 2.24, 1.77, and 0.60 kJy sr$^{-1}$ at K, J and L respectively and are listed in Table~\ref{error} under ``Calibration.''

There is a small correction for faint galaxies that appear in the 2MASS PSC.  These have been subtracted along with the Galactic stars, but should be included in the CIRB.  \citet{elw01} estimates this correction is 0.1 and 0.05 kJy sr$^{-1}$ at 2.2 and 1.25 $\mu$m.  The 0.1 kJy sr$^{-1}$ correction at 2.2 $\mu$m implies a 0.05 kJy sr$^{-1}$ correction at 3.5 $\mu$m due to the relative calibration factor of 0.491.  These corrections have been added back after taking the mean of C = DZ(0) - F in the 40 regions. Thus, the final reported values of the CIRB are the mean of the values C in Tables \ref{ktable}, \ref{jtable} and \ref{ltable} plus this correction.  An uncertainty of 100\% of this correction is included in Table~\ref{error} under ``Galaxies.''

\citet{gor00} adopt an uncertainty of 5\% of the zodiacal intensity at the ecliptic poles.  These uncertainties are listed in Table~\ref{error} under ``Zodiacal.''     

After adding errors in quadrature, we obtain a CIRB of  14.69 $\pm$ 4.49 kJy sr$^{-1}$ at 2.2 $\mu$m, a weak limit of 8.88 $\pm$ 6.26 kJy sr$^{-1}$ at 1.25 $\mu$m and a CIRB of 15.62 $\pm$ 3.34 kJy sr$^{-1}$ at 3.5 $\mu$m.  

\begin{figure}
\epsscale{.80}
\plotone{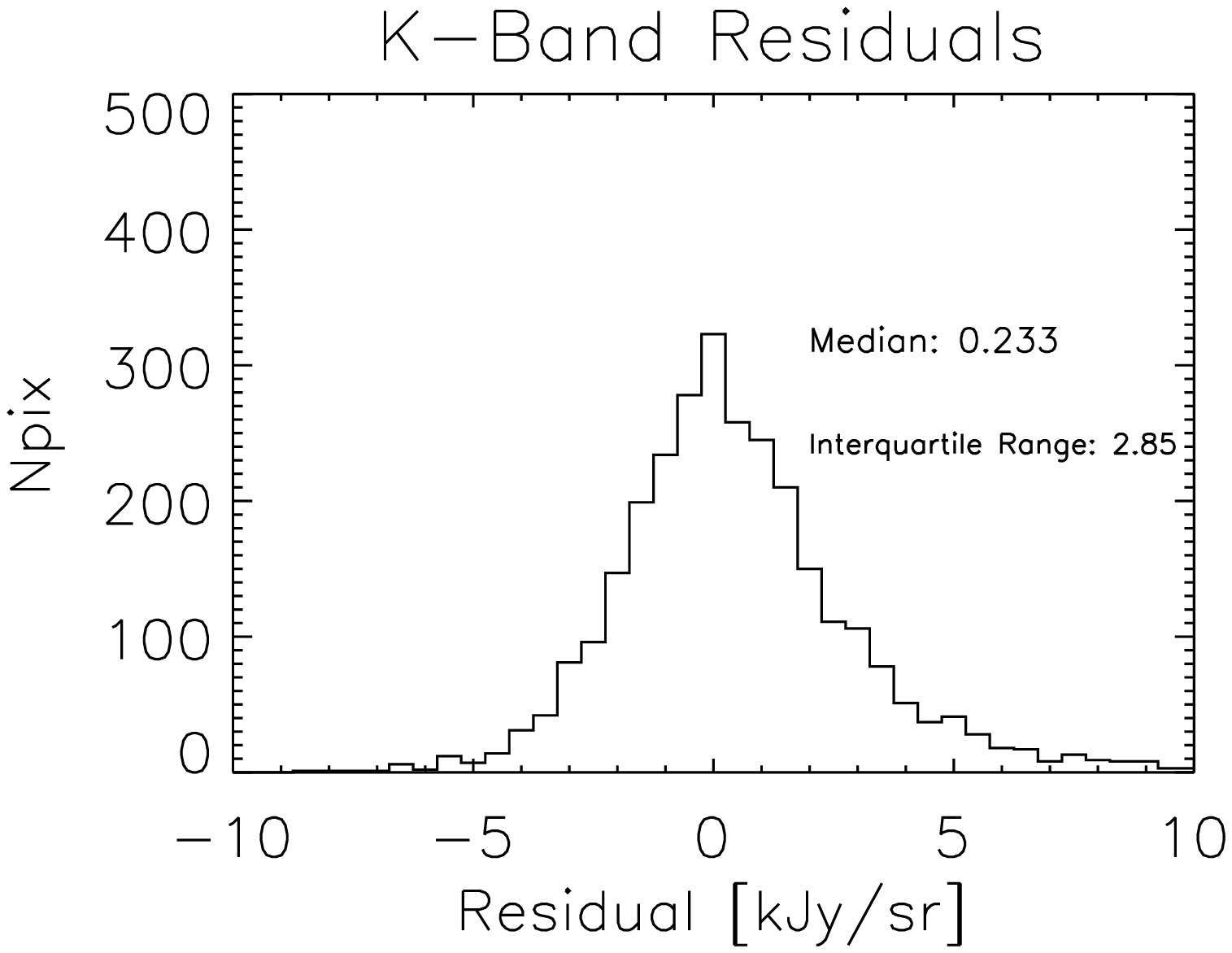}
\caption{Histogram of $DZ_i- \kappa B_i-DZ(0)$ at 2.2 $\mu$m for all regions combined where $\kappa$ = 0.88.} 
\label{kres}
\end{figure}

\begin{figure}
\epsscale{.80}
\plotone{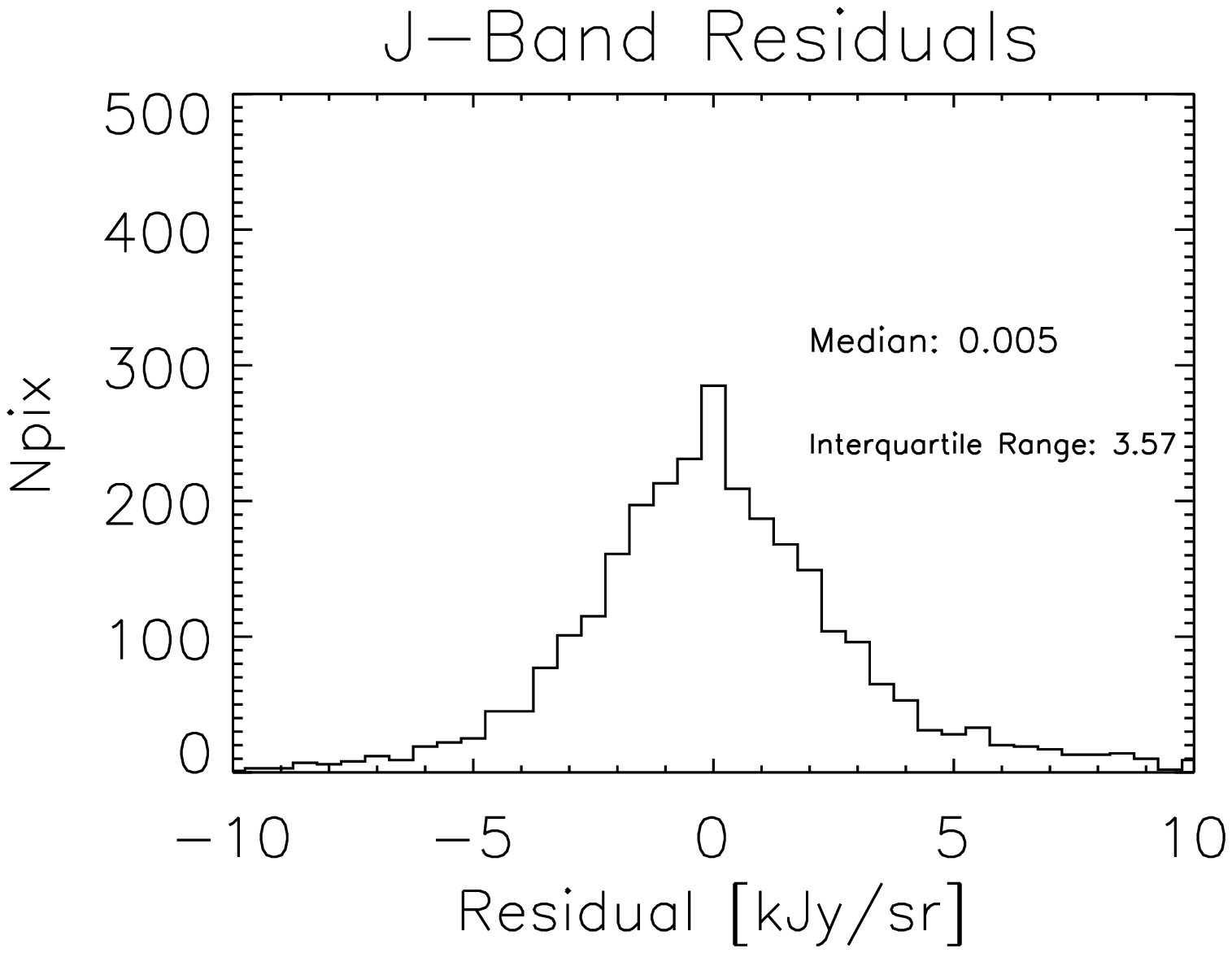}
\caption{Histogram of $DZ_i- \kappa B_i-DZ(0)$ at 1.25 $\mu$m for all regions combined where $\kappa$ = 0.97.} 
\label{jres}
\end{figure}

\begin{figure}
\epsscale{.80}
\plotone{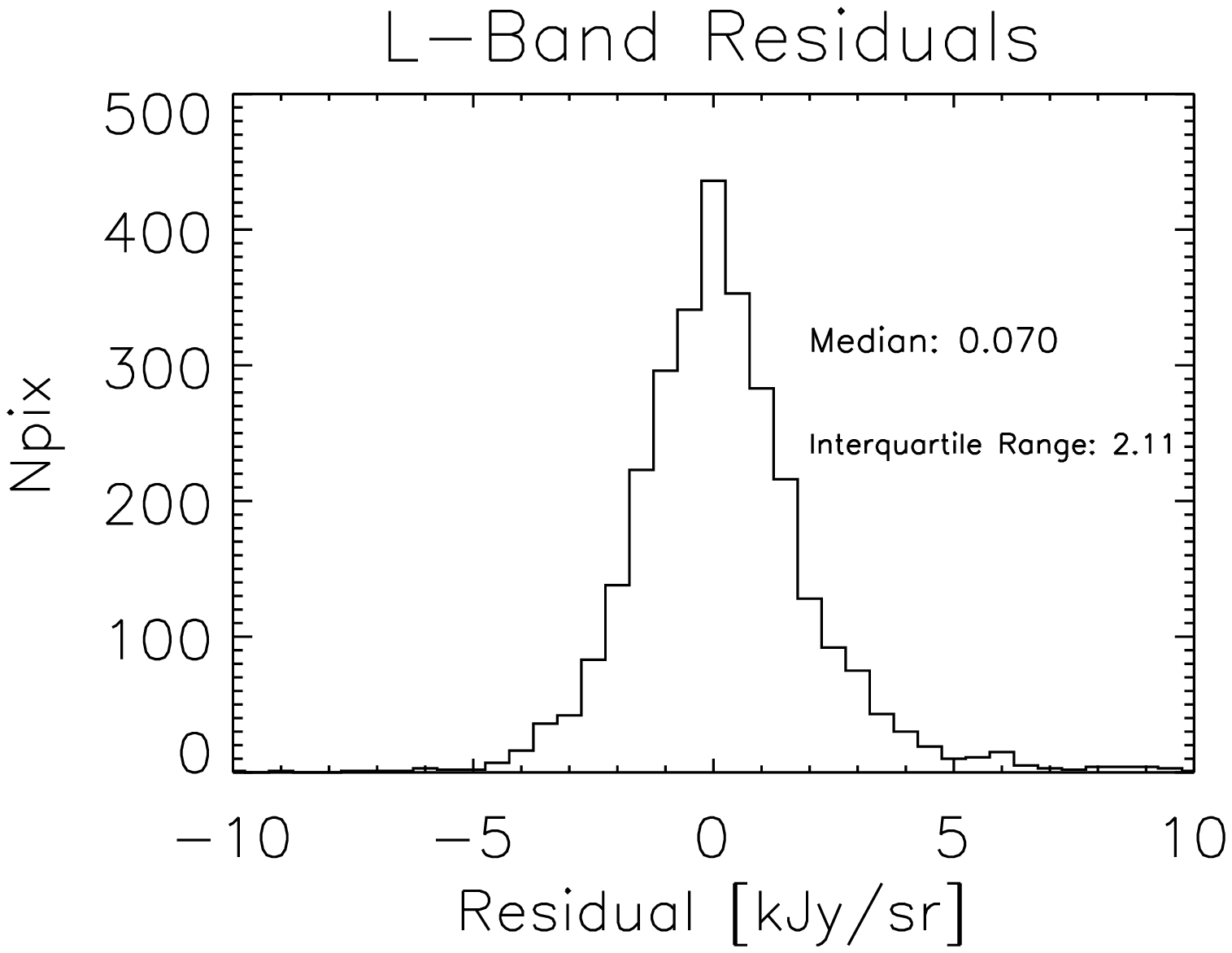}
\caption{Histogram of $DZ_i- \kappa B_i-DZ(0)$ at 3.5 $\mu$m for all regions combined where $\kappa$ = 0.43.} 
\label{lres}
\end{figure}

Figures \ref{kres}, \ref{jres}, and \ref{lres} show histograms of the residuals $DZ_i- \kappa B_i-DZ(0)$ for all 2971 pixels in K, J and L with interquartile ranges of 2.85, 3.57 and 2.11 respectively.  Dividing these residuals by $\sigma(B_i)$ at each pixel gives a non-Gaussian distribution which is tightly packed near zero with a few pixels extending out to large values.    The number of pixels with $|(DZ_i- \kappa B_i-DZ(0))/\sigma(B_i)|$ less than \{0.5, 1.0, 2.0, 3.0\} at K, J and L are \{2424, 2832, 2930, 2951\}, \{2588, 2873, 2955, 2955\} and \{2590, 2867, 2947, 2955\}.  For comparison, with a Gaussian distribution, one would expect these numbers to be, \{1138, 2028, 2835, 2963\}; more spread out in the center, with fewer pixels in the extended tail.  We have used a robust, least sum of absolute values fitting method, which, by it's nature, is insensitive to the few pixels with large residual/$\sigma$.  Thus the fit for the slopes, $\kappa$, and intercepts, DZ(0), is to the pixels in the narrow core, and the outliers have little effect on their derived values and the final CIRB values.  

\begin{figure}
\epsscale{.80}
\plotone{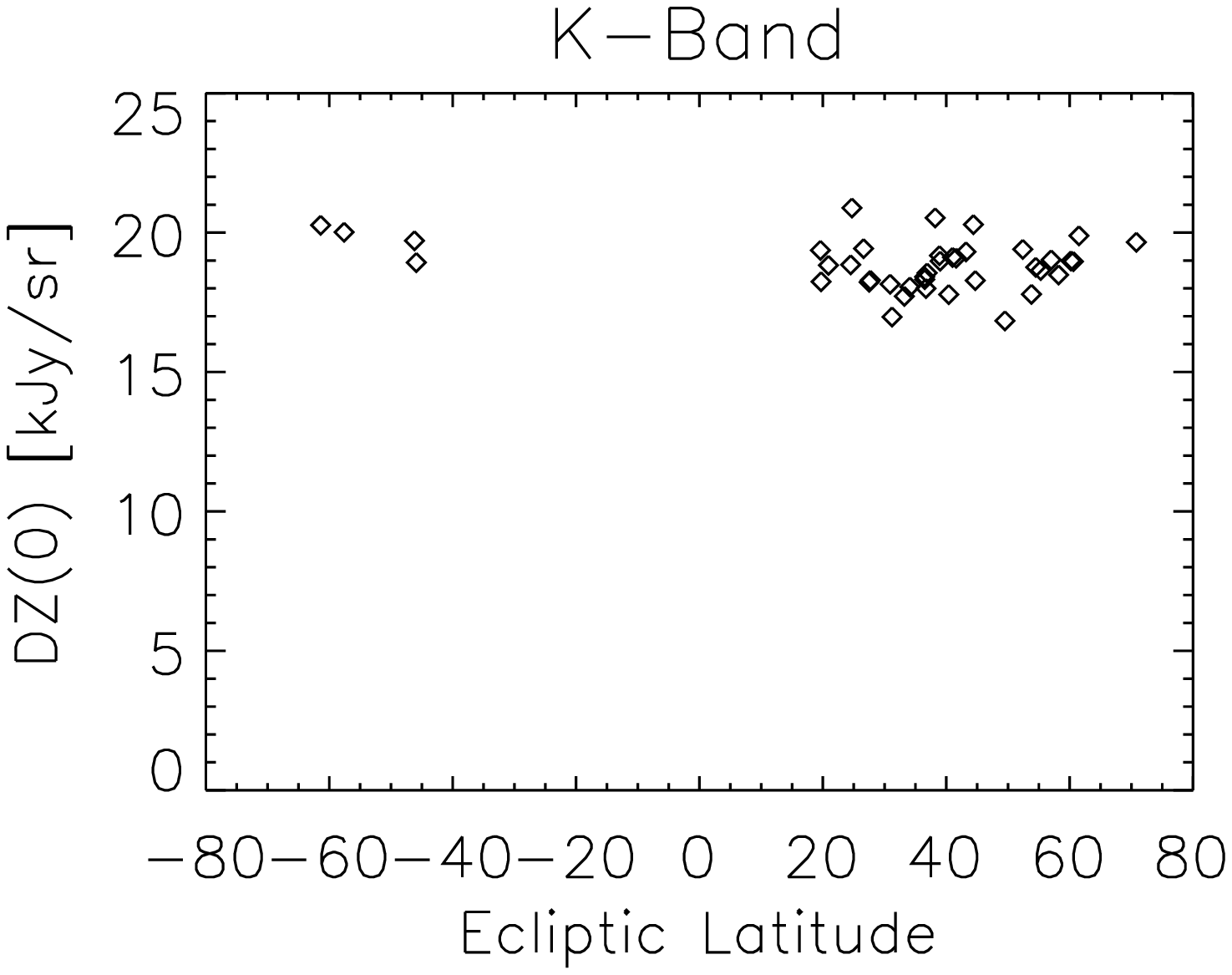}
\caption{Derived intercepts, DZ(0), versus Ecliptic Latitude at 2.2 $\mu$m.} 
\label{kecl}
\end{figure}

\begin{figure}
\epsscale{.80}
\plotone{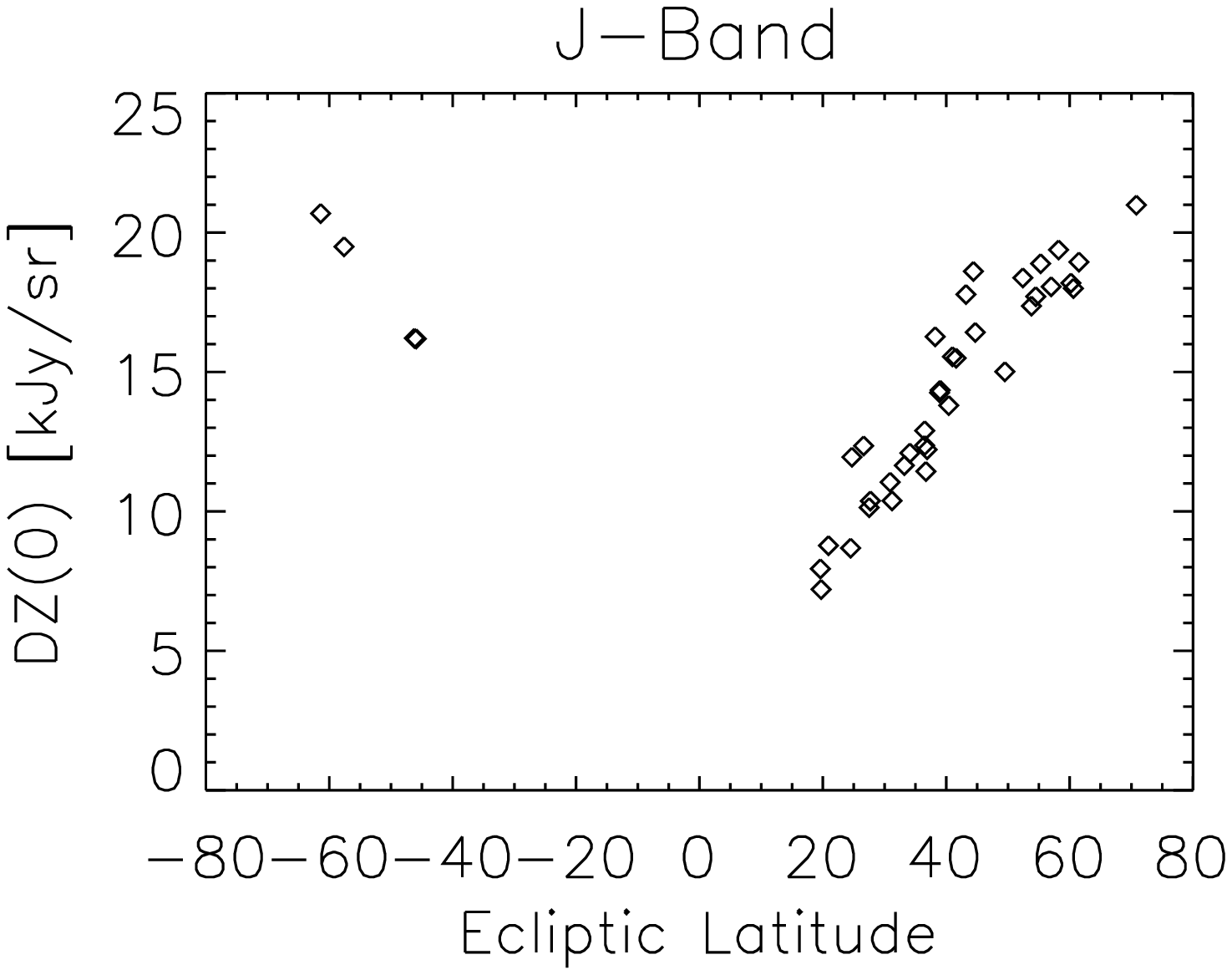}
\caption{Derived intercepts, DZ(0), versus Ecliptic Latitude at 1.25 $\mu$m.} 
\label{jecl}
\end{figure}

\begin{figure}
\epsscale{.80}
\plotone{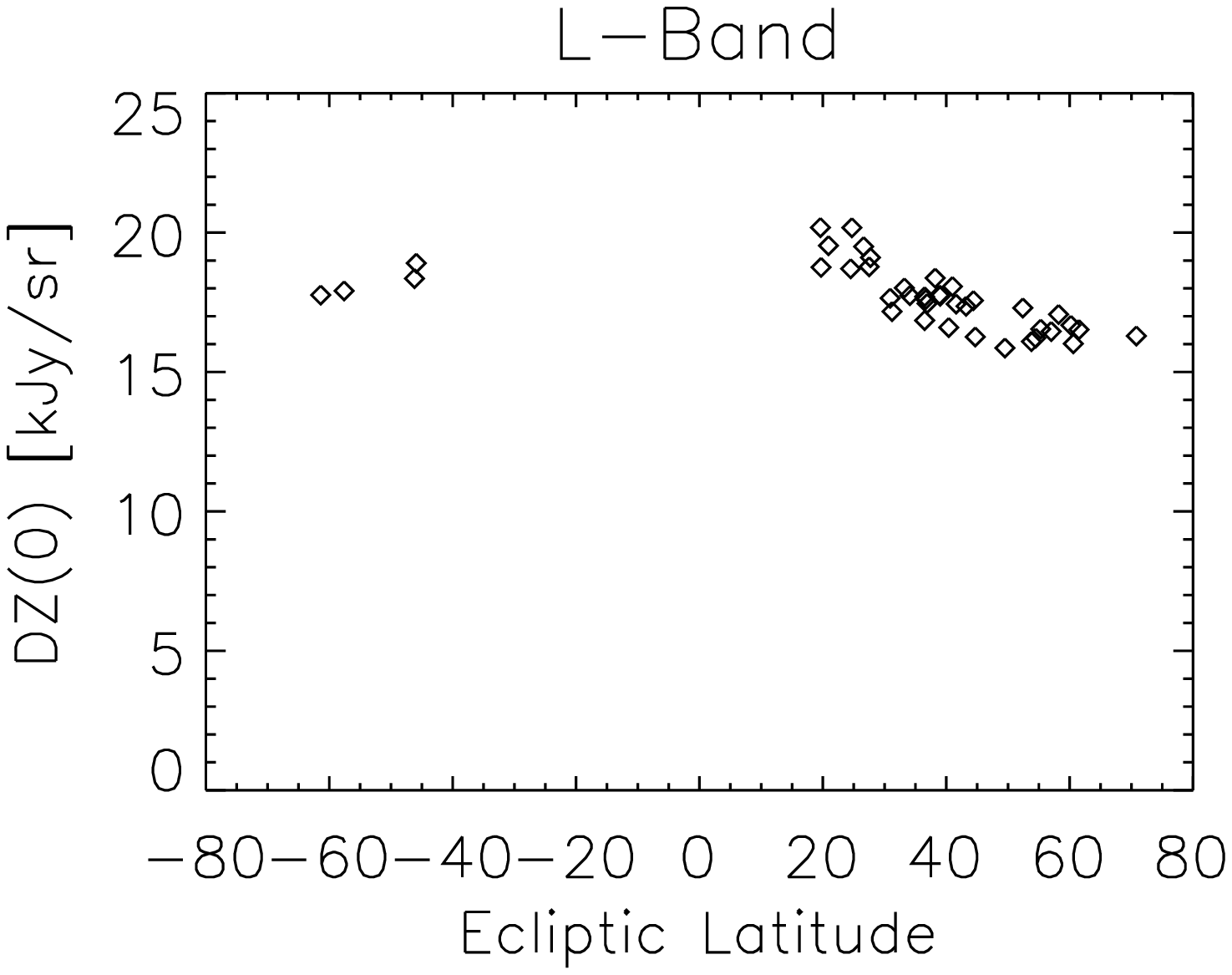}
\caption{Derived intercepts, DZ(0), versus Ecliptic Latitude at 3.5 $\mu$m.} 
\label{lecl}
\end{figure}

Figures \ref{kecl}, \ref{jecl}, and \ref{lecl} show the derived intercepts, DZ(0), vs. ecliptic latitude in the three bands.  We see here the same trends with ecliptic latitude as were apparent in \citet{wrj01}.  While the K-band intercepts appear reasonably independent of ecliptic latitude, there is a strong trend in J and a slight negative trend in the L-band. The zodiacal light is fainter at 3.5 $\mu$m than at 2.2 $\mu$m, and so the stronger dependence on $\beta$ at L than at K may seem surprising.  However, at 3.5 $\mu$m, we begin to see thermal emission from the interplanetary dust, in addition to the scattered sunlight.  From a modeling standpoint, this gives another free parameter which should provide a better fit, but from a physical standpoint, we are likely seeing a more complicated emission/scattering pattern on the sky which is more difficult to model correctly.  There is an overall scaling factor at each wavelength in the zodiacal light model, but the parameters that determine the physical shape of the dust cloud were fit simultaneously to observations at 8 DIRBE bands.  The trend with ecliptic latitude at both J and L indicates a problem with the modeled shape of the cloud. The better fit to the scattered sunlight at K could be a coincidence, rather than better modeling in that band. This remaining solar system dependence in two of the three analyzed bands is evidence of a problem with the zodiacal light model that still prevents us from claiming a detection at 1.25 $\mu$m.  The model may be improved by requiring that these DIRBE minus 2MASS intercepts be ecliptic independent simultaneously in all three bands.  Improvements to the zodiacal light model will be addressed in future work.   

\section{Discussion}

The results of this DIRBE minus 2MASS subtraction in these 40 regions of the sky give a statistically significant isotropic background of  14.69 $\pm$ 4.49 kJy sr$^{-1}$ at 2.2 $\mu$m and 15.62 $\pm$ 3.34 kJy sr$^{-1}$ at 3.5 $\mu$m where the uncertainty has not been significantly reduced since the dominant sources of error are systematic.  These results are consistent with earlier results, summarized in Table~\ref{oldvals}  , including \citet{gor00}, \cite{wrr00}, \citet{elw01} and with the 13 similarly analyzed regions from \citet{wrj01} all of which used the same zodiacal light model considered here \citep{wr98}.

In the J-band, we have weak limit on the 1.25 $\mu$m CIRB of 8.88 $\pm$ 6.26 kJy sr$^{-1}$.  This is also consistent with the 1.25 $\mu$m values  in Table~\ref{oldvals}, which were reported in \cite{elw01} and  \citet{wrj01}, also using the same zodiacal light model used here. The \citet{kel98} zodiacal light model gives a zodiacal intensity at the ecliptic pole 3.9, 9.2 and 4.0 kJy sr$^{-1}$ lower at K, J and L.  Allowing for this difference in the models, these results are also consistent with those reported in \citet{cam01} and \citet{mat00} at 1.25 and 2.2 $\mu$m and \citet{mat05} at 2.2 and 3.5 $\mu$m, also shown in Table~\ref{oldvals}.

The cumulative light from galaxies is a strict lower limit on the CIRB.  A determination of the total contribution of resolved galaxies to the CIRB was determined via galaxy number counts by \citet{faz04}.  Using the InfraRed Array Camera (IRAC) on the Spitzer Space Telescope  \citep{eis04}, surveys were done of the Bo\"{o}tes region, the Extended Groth Strip and a deep image surrounding the QSO HS 1700+6416.  After integrating the light from galaxies from the 10th to the 21st magnitudes, a total integrated intensity of 5.4 nW m$^{-2}$ sr$^{-1}$ (6.5 kJy sr$^{-1}$) at 3.6 $\mu$m is reported, which is less than half of the CIRB determined at that wavelength by this and other similar studies.  This 3.6 $\mu$m value is the final entry in Table~\ref{oldvals}. This discrepancy may be partially resolved by improving the photometry of these and other Spitzer surveys.  But, the dominant source of error in these DIRBE minus 2MASS measurements comes from the estimation of the zodiacal light, and based on the strong ecliptic dependence of our J-band results, it will likely take an improvement of the zodiacal light models to make significant progress in determining whether we can resolve this discrepancy with improvements in data analysis, or if we truly require some exotic diffuse source as suggested by \citet{cam01}.

\begin{figure}
\epsscale{.80}
\plotone{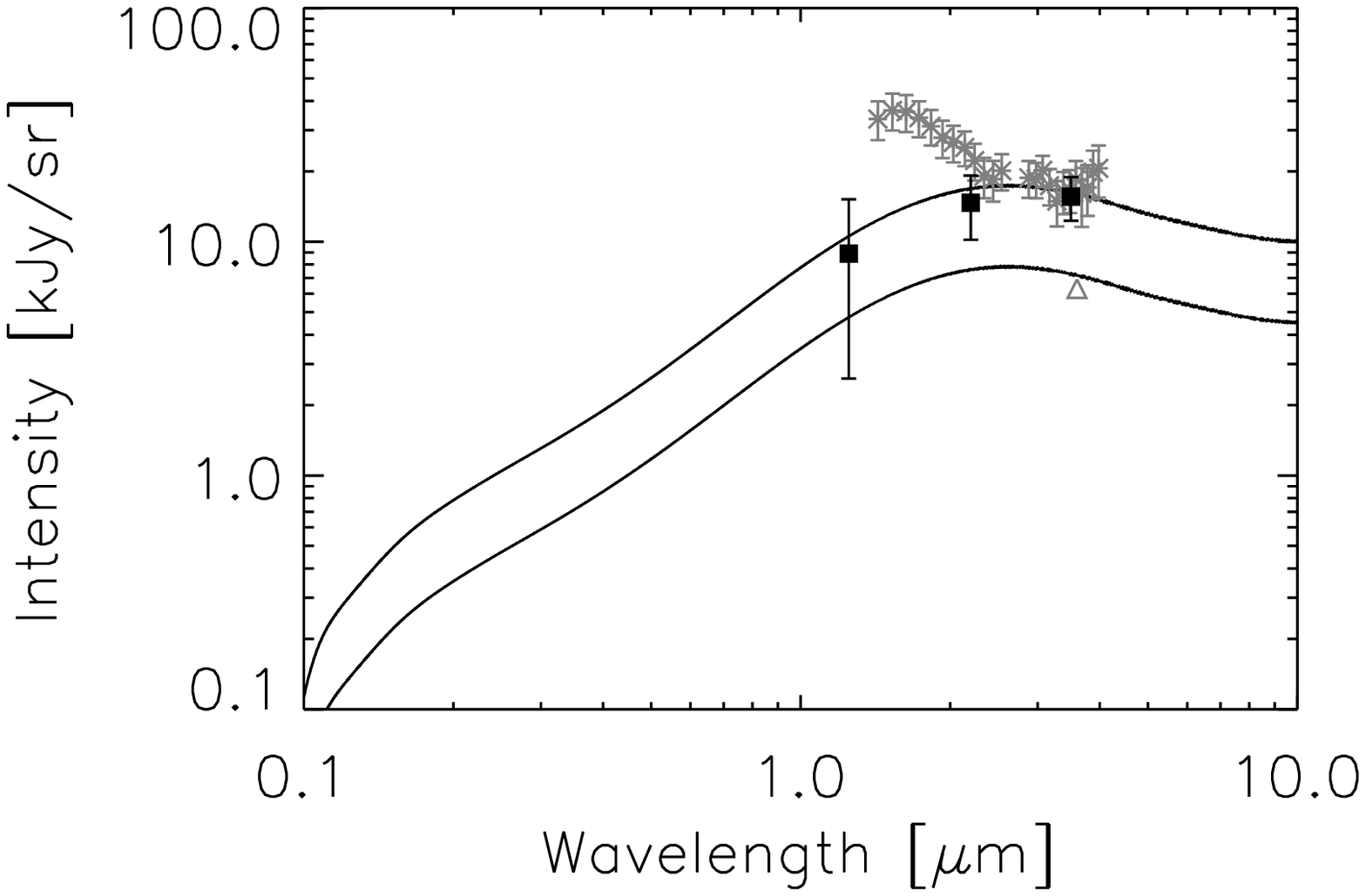}
\caption{Filled black squares are the CIRB values reported here.  Gray stars are \citet{mat05} values from IRTS observations.  The open gray triangle is the \citet{faz04} lower limit at 3.6$\mu$m.  For comparison, the upper and lower black curves are the P1.0 and P0.45 models used by \citet{aha06} to estimate the attenuation of TeV $\gamma$-rays by the CIRB. P1.0 was normalized by \citet{aha06} to fit the 1.25-3.5 $\mu$m values from \citet{dwe98} and \citet{elw01}.  P0.45 is the P1.0 scaled down by a factor of 0.45, which was required, using this shape for the CIRB, to give blazar spectra with power law spectral indices of at least 1.5.} 
\label{p1.0}
\end{figure}

It is also interesting to note another indirect constraint on the CIRB from the attenuation of TeV $\gamma$-rays by $e^+e^-$ pair production through collisions with CIRB photons. A recent attempt by \citet{map04} to fit the TeV spectrum of the blazar H1426+428 found that the spectrum can not be fit using only the integrated light from galaxies.  Their best fit uses the Wright (2001) determination of the CIRB favoring the higher values determined from direct observations of the total sky brightness.  However, H.E.S.S. observations of the blazars H2356-309 and 1ES 1101-232 by \citet{aha06} suggest that the CIRB can not be much higher than the lower limits from galaxy counts.  The \citet{elw01} 1.25 and 2.2 $\mu$m values and the \citet{dwe98} 3.5 $\mu$m value were used by \citet{aha06} in fitting a model of the Extragalactic Background Light (EBL) needed for estimating $\gamma$-ray attenuation, resulting in their P1.0 EBL model which is the top curve in Figure~\ref{p1.0}.  This model gives (11,18,16) kJy sr$^{-1}$ at 1.25, 2.2 and 3.5 $\mu$m and assuming a power law blazar spectrum ($dN/dE$ [photons cm$^{-2}$ s$^{-1}$ TeV$^{-1}$] $\propto$ $E^{-\Gamma}$) results in a power law index for the spectrum of the blazar 1ES 1101-232 of $\Gamma$ = -0.1. The model must be scaled down by a factor of 0.45 to the P0.45 model (lower curve in Figure~\ref{p1.0}) to give a power law index of at least 1.5, which is considered by \citet{aha06} to be the lowest acceptable value.  \citet{map06} show however that an EBL model, also based on the \citet{elw01} values at 1.25 and 2.2 $\mu$m, but with a steeper decline from 4 to 10 $\mu$m results in a power law index of $\Gamma$ = +0.5.  They consider $\Gamma$ = 0.6 to be the lowest acceptable index based on physical considerations and suggest then that while the DIRBE minus 2MASS CIRB values at 1.25 and 2.2 $\mu$m from \citet{elw01} do require a hard spectrum, and the lower limits from galaxy counts are favored, they, and the slightly lower values reported here ((9,15,16) kJy sr$^{-1}$), can not be ruled out based on the current H.E.S.S. data.  While the implications of $\gamma$-ray attenuation for CIRB measurement are still limited by the small number of observed sources, this independent limit on the CIRB will only improve as more blazars are observed by H.E.S.S. and eventually, VERITAS.    

There is still a substantial difference between the \citet{faz04} lower limits from galaxy counts and the intensities determined here. Progress can be made in resolving this discrepancy with improvements in the photometry of survey data from \it Spitzer\rm,  as well as improvements in the zodiacal light models and further data on gamma-ray attenuation. However, the dominant source of error in directly measuring the CIRB, the model based subtraction of the zodiacal light, will not be significantly reduced with currently available data.  A directly measured map of the zodiacal light, which would have to be observed from outside the bulk of the IPD cloud, beyond about 3AU, would allow accurate removal of the zodiacal light from the DIRBE maps and thus an accurate, direct measurement of the Cosmic Infrared Background.  We have shown here that the subtraction of catalogued stars from low resolution maps works well, thus an instrument with a field of view of a few square degrees and a resolution of a few arcminutes would suffice.  The main requirements would be sensitivity to extremely low surface brightnesses, down to less than 1 nW m$^{-2}$ sr$^{-1}$ for good signal to noise, and an accurate, absolute flux calibration.  Such an instrument could be a camera on a probe to one of the outer planets.  It would also be useful to have observations from different positions with respect to the IPD cloud.  This could be accomplished either by observing the same fields at widely different solar elongations during a long lived mission as the craft orbits the sun, or by observing during the cruise from 1 to 3 AU as the dust density decreases.  While we will continue to improve our understanding of the CIRB in the mean time, a space mission of this type will ultimately be required.

\acknowledgments

The COBE datasets were developed by the NASA Goddard Space Flight Center under the direction of the COBE Science Working Group and were provided by the NSSDC.  This publication makes use of the data product from the Two Micron All Sky Survey, which is a joint project of the University of Massachusetts and the Infrared Processing and Analysis Center, funded by the National Aeronautics and Space Administration and the National Science Foundation. This research has also made use of the NASA/ IPAC Infrared Science Archive, which is operated by the Jet Propulsion Laboratory, California Institute of Technology, under contract with the National Aeronautics and Space Administration

\clearpage
\begin{deluxetable}{ccccccc}
\tabletypesize{\scriptsize}
\tablecaption{K-band Results \label{ktable}}
\tablewidth{0pt}
\tablehead{
\colhead{l} & \colhead{b} & \colhead{$\beta$} & \colhead{$N_{pix}$} & \colhead{DZ(0) [kJy sr$^{-1}$]} &\colhead{F [kJy sr$^{-1}$]} &\colhead{C [kJy sr$^{-1}$]} }
\startdata
 204.4    &   67.8    &   20.9    &     79    &  18.83    &   4.11    &  14.72\\
 293.9    &  -72.5    &  -46.2    &     74    &  19.71    &   4.55    &  15.16\\
 191.5    &   72.1    &   26.6    &     75    &  19.43    &   3.83    &  15.60\\
 227.8    &   75.1    &   19.6    &     75    &  19.37    &   3.70    &  15.67\\
 150.5    &   67.6    &   41.0    &     73    &  19.11    &   3.86    &  15.25\\
 193.5    &   76.9    &   27.5    &     75    &  18.23    &   3.68    &  14.55\\
 208.7    &   78.3    &   24.7    &     77    &  20.89    &   3.72    &  17.17\\
 146.7    &   69.3    &   41.6    &     75    &  19.09    &   3.88    &  15.21\\
 150.6    &   72.6    &   38.9    &     72    &  19.18    &   3.90    &  15.28\\
 164.0    &   79.0    &   33.2    &     77    &  17.72    &   3.73    &  13.99\\
 158.3    &   81.1    &   38.2    &     74    &  20.53    &   3.77    &  16.76\\
 127.6    &   53.8    &   57.0    &     76    &  19.03    &   4.93    &  14.10\\
 143.2    &   79.4    &   36.5    &     73    &  18.41    &   3.62    &  14.79\\
 166.0    &   86.3    &   31.2    &     71    &  16.98    &   3.67    &  13.31\\
 133.5    &   80.8    &   36.7    &     73    &  18.00    &   3.63    &  14.37\\
 317.5    &   76.8    &   19.7    &     74    &  18.24    &   3.96    &  14.28\\
 115.4    &   65.7    &   52.4    &     74    &  19.40    &   4.18    &  15.22\\
 355.1    &   81.7    &   27.7    &     75    &  18.29    &   4.07    &  14.22\\
  43.5    &   82.9    &   34.1    &     73    &  18.05    &   3.88    &  14.17\\
 115.8    &   56.4    &   60.2    &     76    &  18.99    &   4.74    &  14.25\\
  72.2    &   81.9    &   36.9    &     73    &  18.55    &   3.89    &  14.66\\
  15.3    &   81.9    &   30.9    &     77    &  18.15    &   3.85    &  14.30\\
  79.3    &   79.1    &   40.4    &     72    &  17.78    &   3.90    &  13.88\\
 347.9    &   76.1    &   24.5    &     74    &  18.85    &   4.10    &  14.75\\
  49.7    &   79.4    &   36.5    &     74    &  18.32    &   4.67    &  13.65\\
 111.6    &   57.0    &   60.6    &     75    &  18.97    &   4.61    &  14.36\\
  93.3    &   70.0    &   49.5    &     72    &  16.84    &   4.02    &  12.82\\
 103.0    &   57.8    &   61.5    &     73    &  19.89    &   4.80    &  15.09\\
  42.2    &   72.6    &   39.0    &     72    &  18.98    &   4.23    &  14.75\\
  56.4    &   68.5    &   44.7    &     75    &  18.28    &   4.25    &  14.03\\
  83.8    &   63.3    &   55.3    &     73    &  18.65    &   4.44    &  14.21\\
 109.1    &   46.2    &   70.8    &     73    &  19.66    &   5.75    &  13.91\\
  84.7    &   60.4    &   58.2    &     75    &  18.48    &   4.73    &  13.75\\
  74.3    &   62.1    &   54.5    &     76    &  18.76    &   4.52    &  14.24\\
  44.6    &   64.0    &   43.2    &     76    &  19.31    &   4.61    &  14.70\\
  69.8    &   61.2    &   53.8    &     76    &  17.79    &   4.82    &  12.97\\
  45.1    &   59.8    &   44.4    &     71    &  20.29    &   5.07    &  15.22\\
 252.6    &  -71.8    &  -45.9    &     74    &  18.93    &   4.42    &  14.51\\
 255.8    &  -59.2    &  -57.6    &     77    &  20.02    &   5.13    &  14.89\\
 254.1    &  -53.0    &  -61.4    &     74    &  20.27    &   5.60    &  14.67\\
\enddata

\tablecomments{Mean of the CIRB over all 40 regions (\bf 2971 \rm pixels) at 2.2 microns is \bf 14.59 $\pm$ 0.14 kJy sr$^{-1}$ \rm.  This uncertainty is listed in Table~\ref{error} under ``Scatter.''}

\tablenotetext{a}{Intercepts derived using a weighted median procedure resulting in a global slope of 0.88 and 40 separate intercepts, DZ(0).}

\end{deluxetable}

\clearpage

\begin{deluxetable}{ccccccc}
\tabletypesize{\scriptsize}
\tablecaption{J-band Results \label{jtable}}
\tablewidth{0pt}
\tablehead{
\colhead{l} & \colhead{b} & \colhead{$\beta$} & \colhead{$N_{pix}$} & \colhead{DZ(0)\tablenotemark{b} [kJy sr$^{-1}$]} &\colhead{F [kJy sr$^{-1}$]} &\colhead{C [kJy sr$^{-1}$]} }
\startdata
 204.4    &   67.8    &   20.9    &     79    &   8.77    &   5.60    &   3.17\\
 293.9    &  -72.5    &  -46.2    &     74    &  16.21    &   6.20    &  10.01\\
 191.5    &   72.1    &   26.6    &     75    &  12.35    &   5.38    &   6.97\\
 227.8    &   75.1    &   19.6    &     75    &   7.94    &   5.12    &   2.82\\
 150.5    &   67.6    &   41.0    &     73    &  15.55    &   5.32    &  10.23\\
 193.5    &   76.9    &   27.5    &     75    &  10.14    &   4.92    &   5.22\\
 208.7    &   78.3    &   24.7    &     77    &  11.95    &   5.11    &   6.84\\
 146.7    &   69.3    &   41.6    &     75    &  15.50    &   5.40    &  10.10\\
 150.6    &   72.6    &   38.9    &     72    &  14.26    &   5.29    &   8.97\\
 164.0    &   79.0    &   33.2    &     77    &  11.65    &   5.14    &   6.51\\
 158.3    &   81.1    &   38.2    &     74    &  16.27    &   5.17    &  11.10\\
 127.6    &   53.8    &   57.0    &     76    &  18.06    &   6.74    &  11.32\\
 143.2    &   79.4    &   36.5    &     73    &  12.90    &   4.95    &   7.95\\
 166.0    &   86.3    &   31.2    &     71    &  10.38    &   5.01    &   5.37\\
 133.5    &   80.8    &   36.7    &     73    &  11.44    &   4.89    &   6.55\\
 317.5    &   76.8    &   19.7    &     74    &   7.20    &   5.45    &   1.75\\
 115.4    &   65.7    &   52.4    &     74    &  18.38    &   5.81    &  12.57\\
 355.1    &   81.7    &   27.7    &     75    &  10.38    &   5.51    &   4.87\\
  43.5    &   82.9    &   34.1    &     73    &  12.10    &   5.41    &   6.69\\
 115.8    &   56.4    &   60.2    &     76    &  18.19    &   6.44    &  11.75\\
  72.2    &   81.9    &   36.9    &     73    &  12.22    &   5.30    &   6.92\\
  15.3    &   81.9    &   30.9    &     77    &  11.05    &   5.14    &   5.91\\
  79.3    &   79.1    &   40.4    &     72    &  13.80    &   5.11    &   8.69\\
 347.9    &   76.1    &   24.5    &     74    &   8.69    &   5.54    &   3.15\\
  49.7    &   79.4    &   36.5    &     74    &  12.37    &   6.50    &   5.87\\
 111.6    &   57.0    &   60.6    &     75    &  18.00    &   6.23    &  11.77\\
  93.3    &   70.0    &   49.5    &     72    &  15.01    &   5.56    &   9.45\\
 103.0    &   57.8    &   61.5    &     73    &  18.95    &   6.57    &  12.38\\
  42.2    &   72.6    &   39.0    &     72    &  14.35    &   5.97    &   8.38\\
  56.4    &   68.5    &   44.7    &     75    &  16.42    &   5.82    &  10.60\\
  83.8    &   63.3    &   55.3    &     73    &  18.89    &   5.98    &  12.91\\
 109.1    &   46.2    &   70.8    &     73    &  20.99    &   8.11    &  12.88\\
  84.7    &   60.4    &   58.2    &     75    &  19.38    &   6.43    &  12.95\\
  74.3    &   62.1    &   54.5    &     76    &  17.71    &   6.28    &  11.43\\
  44.6    &   64.0    &   43.2    &     76    &  17.78    &   6.42    &  11.36\\
  69.8    &   61.2    &   53.8    &     76    &  17.37    &   6.64    &  10.73\\
  45.1    &   59.8    &   44.4    &     71    &  18.61    &   6.91    &  11.70\\
 252.6    &  -71.8    &  -45.9    &     74    &  16.19    &   6.12    &  10.07\\
 255.8    &  -59.2    &  -57.6    &     77    &  19.50    &   7.05    &  12.45\\
 254.1    &  -53.0    &  -61.4    &     74    &  20.69    &   7.76    &  12.93\\
\enddata

\tablecomments{Mean of the CIRB over all 40 regions (\bf 2971 \rm pixels) at 1.25 microns is \bf 8.83 $\pm$ 0.51 kJy sr$^{-1}$ \rm.  This uncertainty is listed in Table~\ref{error} under ``Scatter.''}

\tablenotetext{b}{Intercepts derived using a weighted median procedure resulting in a global slope of 0.97 and 40 separate intercepts, DZ(0).}

\end{deluxetable}

\clearpage

\begin{deluxetable}{ccccccc}
\tabletypesize{\scriptsize}
\tablecaption{L-band Results \label{ltable}}
\tablewidth{0pt}
\tablehead{
\colhead{l} & \colhead{b} & \colhead{$\beta$} & \colhead{$N_{pix}$} & \colhead{DZ(0)\tablenotemark{c} [kJy sr$^{-1}$]} &\colhead{F [kJy sr$^{-1}$]} &\colhead{C [kJy sr$^{-1}$]} }
\startdata
 204.4    &   67.8    &   20.9    &     79    &  19.54    &   2.00    &  17.54\\
 293.9    &  -72.5    &  -46.2    &     74    &  18.35    &   2.21    &  16.14\\
 191.5    &   72.1    &   26.6    &     75    &  19.50    &   1.86    &  17.64\\
 227.8    &   75.1    &   19.6    &     75    &  20.18    &   1.80    &  18.38\\
 150.5    &   67.6    &   41.0    &     73    &  18.07    &   1.88    &  16.19\\
 193.5    &   76.9    &   27.5    &     75    &  18.78    &   1.79    &  16.99\\
 208.7    &   78.3    &   24.7    &     77    &  20.18    &   1.81    &  18.37\\
 146.7    &   69.3    &   41.6    &     75    &  17.44    &   1.89    &  15.55\\
 150.6    &   72.6    &   38.9    &     72    &  17.77    &   1.90    &  15.87\\
 164.0    &   79.0    &   33.2    &     77    &  18.02    &   1.81    &  16.21\\
 158.3    &   81.1    &   38.2    &     74    &  18.37    &   1.84    &  16.53\\
 127.6    &   53.8    &   57.0    &     76    &  16.46    &   2.40    &  14.06\\
 143.2    &   79.4    &   36.5    &     73    &  17.70    &   1.76    &  15.94\\
 166.0    &   86.3    &   31.2    &     71    &  17.17    &   1.79    &  15.38\\
 133.5    &   80.8    &   36.7    &     73    &  17.60    &   1.77    &  15.83\\
 317.5    &   76.8    &   19.7    &     74    &  18.76    &   1.93    &  16.83\\
 115.4    &   65.7    &   52.4    &     74    &  17.30    &   2.03    &  15.27\\
 355.1    &   81.7    &   27.7    &     75    &  19.11    &   1.98    &  17.13\\
  43.5    &   82.9    &   34.1    &     73    &  17.74    &   1.89    &  15.85\\
 115.8    &   56.4    &   60.2    &     76    &  16.68    &   2.31    &  14.37\\
  72.2    &   81.9    &   36.9    &     73    &  17.46    &   1.90    &  15.56\\
  15.3    &   81.9    &   30.9    &     77    &  17.65    &   1.87    &  15.78\\
  79.3    &   79.1    &   40.4    &     72    &  16.60    &   1.90    &  14.70\\
 347.9    &   76.1    &   24.5    &     74    &  18.71    &   1.99    &  16.72\\
  49.7    &   79.4    &   36.5    &     74    &  16.85    &   2.27    &  14.58\\
 111.6    &   57.0    &   60.6    &     75    &  16.02    &   2.24    &  13.78\\
  93.3    &   70.0    &   49.5    &     72    &  15.86    &   1.96    &  13.90\\
 103.0    &   57.8    &   61.5    &     73    &  16.52    &   2.34    &  14.18\\
  42.2    &   72.6    &   39.0    &     72    &  17.72    &   2.06    &  15.66\\
  56.4    &   68.5    &   44.7    &     75    &  16.26    &   2.07    &  14.19\\
  83.8    &   63.3    &   55.3    &     73    &  16.54    &   2.16    &  14.38\\
 109.1    &   46.2    &   70.8    &     73    &  16.30    &   2.80    &  13.50\\
  84.7    &   60.4    &   58.2    &     75    &  17.07    &   2.30    &  14.77\\
  74.3    &   62.1    &   54.5    &     76    &  16.22    &   2.20    &  14.02\\
  44.6    &   64.0    &   43.2    &     76    &  17.36    &   2.24    &  15.12\\
  69.8    &   61.2    &   53.8    &     76    &  16.09    &   2.34    &  13.75\\
  45.1    &   59.8    &   44.4    &     71    &  17.57    &   2.47    &  15.10\\
 252.6    &  -71.8    &  -45.9    &     74    &  18.90    &   2.15    &  16.75\\
 255.8    &  -59.2    &  -57.6    &     77    &  17.91    &   2.50    &  15.41\\
 254.1    &  -53.0    &  -61.4    &     74    &  17.76    &   2.73    &  15.03\\
\enddata

\tablecomments{Mean of the CIRB over all 40 regions (\bf 2971 \rm pixels) at 3.5 microns is \bf 15.57 $\pm$ 0.20 kJy sr$^{-1}$ \rm.  This standard deviation of the mean is listed in Table~\ref{error} under ``Scatter.''}

\tablenotetext{c}{Intercepts derived using a weighted median procedure resulting in a global slope of 0.43 and 40 separate intercepts, DZ(0).}

\end{deluxetable}

\begin{deluxetable}{lcccl}
\tablecaption{Previous determinations of the CIRB [kJy sr$^{-1}$]\label{oldvals}}
\tablewidth{0pt}
\tablehead{
\colhead{Authors} & \colhead{1.25 $\mu$m} & \colhead{2.2 $\mu$m} & \colhead{3.5 $\mu$m} & \colhead{Model} }
\startdata
\citet{gor00}	& \nodata & 16.4 $\pm $ 4.4 & 12.8 $\pm$ 3.8& \citet{wr98}\\
\citet{wrr00}	& \nodata & 16.9 $\pm $4.4 & 14.4 $\pm $3.7 & \citet{wr98}\\
\citet{elw01}	& 12 $\pm$7 & 14.8 $\pm$ 4.6 & \nodata & \citet{wr98}\\
\citet{wrj01}	& 10.1 $\pm$ 7.4 & 17.6 $\pm$ 4.4 & 16.1 $\pm$ 4 & \citet{wr98}\\
\it This Work \rm	& \it 8.9 $\pm $6.3 \rm & \it 14.7 $\pm$ 4.5 \rm & \it 15.6 $\pm$ 3.3\rm &\it \citet{wr98}\rm\\
\citet{mat00}	& 25 $\pm$ 6.3	& 20.5 $\pm $ 3.7 	& \nodata & \citet{kel98}	\\
\citet{cam01}	& 22.9 $\pm$ 7.0 & 20.4 $\pm $ 4.9 & \nodata & \citet{kel98}\\
\citet{mat05}	& \nodata & 22.3 $\pm $ 4 	& 16.9 $\pm $ 3.5 & \citet{kel98}	\\
\citet{faz04}	& \nodata & \nodata & $>$ 6.5 & N.A. (galaxy counts)\\
\enddata

\tablecomments{The \citet{kel98} model gives a CIRB 9.2, 3.9 \& 4.0 kJy sr$^{-1}$ higher at 1.25, 2.2 and 3.5 $\mu$m.}
\end{deluxetable}

\clearpage

\begin{deluxetable}{lccc}
\tablecaption{Error Budget for the CIRB \label{error}}
\tablewidth{0pt}
\tablehead{
\colhead{Component} & \colhead{1.25 $\mu$m} & \colhead{2.2 $\mu$m} & \colhead{3.5 $\mu$m} }
\startdata
Scatter		& 0.51 & 0.14 & 0.20\\
Faint Source	& 1.17 & 0.85 & 0.42\\
Galaxies	& 0.05 & 0.10 & 0.05\\
Calibration	& 1.77 & 2.24 & 0.60\\
Zodiacal	& 5.87 & 3.79 & 3.25\\
Quadrature Sum	& 6.26 & 4.49 & 3.34\\
\enddata
\end{deluxetable}

\clearpage

\clearpage

\clearpage

\clearpage

\clearpage

\clearpage

\clearpage

\clearpage

\clearpage

\clearpage

\end{document}